\documentclass{article}
\pdfoutput=1
\usepackage[onehalfspacing]{setspace}
\usepackage{fullpage}
\usepackage{sectsty} \sectionfont{\large}
\usepackage{amsmath,amsfonts,bm}
\usepackage{url,graphicx,natbib} 

\DeclareMathOperator*{\var}{var}
\DeclareMathOperator*{\cov}{cov}
\DeclareMathOperator*{\corr}{corr}
\newcommand{\R}{\mathbb{R}}
\newcommand{\bpi}{{\bm\pi}}
\newcommand{\brho}{{\bm\rho}}

\def\svndate$#1: #2-#3-#4 #5 #6 (#7) ${\def\fmtdate{#2--#3--#4}}
\svndate$Date: 2010-11-16 16:48:37 -0500 (Tue, 16 Nov 2010) $

\begin{document}

\begin{center}

{\small Technical Report No.\ 1010,
 Department of Statistics, University of Toronto}

\vspace*{0.6in}

{\Large\bf Graphical Comparison of MCMC Performance} \\[16pt]

{\large Madeleine B. Thompson}\\[3pt]
 \url{mthompson@utstat.toronto.edu} \\
 \url{http://www.utstat.toronto.edu/mthompson} \\
November 16, 2010
\end{center}

\vspace*{6pt}

\bigskip

\begin{abstract}
This paper presents a graphical method for comparing performance
of Markov Chain Monte Carlo methods.  Most researchers present
comparisons of MCMC methods using tables of figures of merit; this
paper presents a graphical alternative.  It first discusses the
computation of autocorrelation time, then uses this to construct a
figure of merit, log density function evaluations per independent
observation.  Then, it demonstrates how one can plot this figure
of merit against a tuning parameter in a grid of plots where columns
represent sampling methods and rows represent distributions.  This
type of visualization makes it possible to convey a greater depth
of information without overwhelming the user with numbers, allowing
researchers to put their contributions into a broader context than
is possible with a textual presentation.
\end{abstract}

\section{Introduction}

Markov Chain Monte Carlo is a class of computational methods with which
a user simulates a Markov chain whose stationary distribution is a
target distribution of interest, generating a dependent sample from
the target distribution.  When choosing between MCMC methods, a
user must trade off their own effort, computation time, and accuracy
of results.

While many methods are documented in the literature, end-users
gravitate to two classic methods: Gibbs sampling (when the full
conditional distributions can be sampled from) and Metropolis--Hastings
updates with Gaussian proposals (when they cannot).  These are
widely available and well understood, but often inefficient.  Perhaps,
if better comparisons of MCMC methods were possible, a consensus
would emerge on which widely applicable modern methods perform best.

In the computational statistics literature, most researchers
compare the computational efficiency of methods with tables of
autocorrelation times, effective sample sizes, or other similar
measures.  For example, in the Journal of Computational and Graphical
Statistics Vol.~19 No.~2, four papers presented comparisons of MCMC
methods; all four compared their proposed methods this way
\citep{giordani10,liechty10,niemi10,papastamoulis10}.

By using tabular comparisons, researchers limit themselves to
presenting only a few data points.  This is compelling when the
target distribution is important in its own right, the set of
reasonable samplers is small, and none of the samplers require
tuning.  However, if a researcher wants to present a method for
general use, readers will find a broader comparison useful.  This
paper presents a graphical alternative to tables
that allows researchers to compare their MCMC methods to existing
ones without overwhelming readers with numbers.

To quantify the efficiency of a simulation, I will use the
\textit{autocorrelation time}---also called correlation length---the
number of observations from an MCMC simulation equivalent to a
single independent observation.  If an MCMC simulation has an
autocorrelation time of $\tau$, estimates based on this simulation
will have standard errors approximately the same as those of an independent
sample a factor of $\tau$ smaller.  Section~\ref{section-corlen}
will discuss autocorrelation time in detail.

Section~\ref{section-example} describes a collection of four
distributions and four MCMC methods.  Using these distributions and
methods, section~\ref{section-cpu} shows that measuring a method's
speed with either processor usage or log-density function evaluations
will often lead to the same conclusions.

Section~\ref{section-results} presents the main contribution of
this paper: a method for comparing MCMC methods by plotting log
density evaluations per iteration times autocorrelation time against a tuning
parameter in a grid of plots where rows represent distributions and
columns represent methods.  This is supported with an example using
the distributions and methods described in section~\ref{section-example}.

Section~\ref{second-demo} develops this idea further, showing that
any three dimensions of the comparison scenario may be examined
simultaneously.  This is demonstrated with an example that shows
how a single method's performance varies as a function of problem
dimension and two tuning parameters.

\section{Autocorrelation time}
\label{section-corlen}

Suppose the goal in running an MCMC simulation is to estimate the
expectation of a random variable $X$ that is some function of the
Markov chain state.  Based on a sample $(X_1,\ldots,X_n)$, we can
estimate $E(X)$ with:
\begin{equation}
\hat\mu_n = \frac{X_1 + \cdots + X_n}{n}
\end{equation}
Assuming $X$ has finite variance, define the autocorrelation function
(ACF) at lag $k$ as:
\begin{equation}\label{rhok}
\rho_k = \frac{\cov(X_i,X_{i+k})}{\var(X_i)}
\end{equation}
Assume the chain has reached its stationary distribution, so
$\rho_k$ does not depend on $i$.  The autocorrelation time is then
\citep[p.~91]{straatsma86}:
\begin{equation}\label{corlen}
\tau = 1 + 2 \sum_{k=1}^{\infty} \rho_k
\end{equation}
If this sum converges---which it will for all geometrically ergodic
Markov chains \citep{mira98}---then:
\begin{equation}\label{meanvar}
\frac{n}\tau \var(\hat\mu_n) \rightarrow \var(X_i)
\end{equation}
When the states are independent, the autocorrelations are zero, so
according to equation~\ref{corlen}, $\tau$ is one, and
equation~\ref{meanvar} becomes the usual formula for the variance of a
sample mean.

\subsection{Estimating the autocorrelation time}
\label{estimating}

One cannot estimate the autocorrelation time by plugging sample
autocorrelations into equation~\ref{corlen}.  The usual formula for the
sample ACF (based on equation~\ref{rhok}) is:
\begin{equation}\label{sampleacf}
\hat\rho_k = \frac1{ns_X^2} \sum_{i=1}^{n-k} (X_i-\hat\mu_n)(X_{i+k}-\hat\mu_n)
\end{equation}
where $s_X^2$ is the sample variance of $(X_1,\ldots,X_n)$.  The
sample ACF is not defined for lags greater than $n-1$, and $\hat\rho_1
+ \cdots + \hat\rho_{n-1}$, the sum of the first $n-1$ terms of
equation~\ref{corlen}, has a variance that does not converge to
zero as the sample size goes to infinity, so an estimator of $\tau$
based on this sum would not be consistent.

Instead, taking the approach of CODA's \texttt{spectrum0.ar} and
\texttt{effectiveSize} functions \citep{plummer06}, we model the
Markov chain as an AR($p$) process, where $p$ is chosen with AIC.
Because the ACF of an AR($p$) process is a function
of its model parameters, we can avoid explicitly computing
the ACF for large lags.  Suppose we have an AR($p$) model of the form:
\begin{equation}\label{armodel}
X_t = \mu + \pi_1 X_{t-1} + \cdots + \pi_p X_{t-p} + a_t,
\qquad a_t \sim N(0,\sigma_a^2)
\end{equation}
Let $\bpi$ denote the vector $(\pi_1,\ldots,\pi_p)^T$.  Let $\brho$
denote the vector $(\rho_1,\ldots,\rho_p)^T$, with $\hat\brho$
denoting its sample estimate from equation~\ref{sampleacf}.  The
true mean, $\mu$, is often known for test distributions; if it is
not, we can estimate it with $\hat\mu_n$.  We use these estimates
to compute an estimate of $\bpi$, denoted by $\hat\bpi$, and the
asymptotic covariance matrix of $\hat\bpi$, denoted by ${\bm V}$, with the
Yule--Walker method \citep[pp.~136--138]{wei06}.

Denote a vector of $p$ ones by ${\bm 1}_p$.  The autocorrelation time of
the process defined by equation~\ref{armodel} is: 
\begin{equation}\label{corlen-ar}
\tau = \frac{1-\brho^T \bpi}{\left(1-{\bm 1}_p^T \bpi\right)^2}
\end{equation}
By substituting $\hat\brho$ for $\brho$ and $\hat\bpi$ for $\bpi$ in
equation~\ref{corlen-ar}, we obtain an estimate of the
autocorrelation time:
\begin{equation}\label{est-corlen}
\hat\tau = \frac{1-\hat\brho^T \hat\bpi}{\left(1-{\bm 1}_p^T \hat\bpi\right)^2}
\end{equation}
Equation~\ref{corlen-ar} is based on the observation
of \citet{heidelberger81} that the autocorrelation time and the
spectrum of a time series at frequency zero are identical.
The spectrum of an autoregressive process is defined by equation
12.2.8b of \citet[pp.~274--275]{wei06}.  I use equation 7.1.5 of
\citet[p.~137]{wei06} to remove the dependence of Wei's equation
12.2.8b on the innovation variance, $\sigma_a^2$.  For more information
on modeling simulations with AR processes, see \citet[\S5.10]{fishman78}.

\subsection{Confidence intervals}

A measure of the uncertainty in the estimated autocorrelation time
is often useful.  Here, I describe one method for generating
confidence intervals for $\tau$ using ${\bm V}$, the asymptotic covariance
matrix of $\hat\bpi$.

The sample ACF to lag $p$ is a function of $\hat\bpi$, so
equation~\ref{est-corlen} does not depend monotonically on $\hat\bpi$
as one might think, making deriving an expression for a closed-form
confidence interval for $\tau$ difficult.  Instead, one can simulate
a set $(\bpi^{(1)}, \ldots, \bpi^{(m)})$ with elements that are each
multivariate Gaussian with mean $\hat\bpi$ and covariance ${\bm V}$.  Each
$\bpi^{(k)}$ can be used to generate a corresponding $\brho^{(k)}$
\citep[p.~47, equation~3.1.28]{wei06} as done in the R function
\texttt{ARMAacf}.  These $\bpi^{(k)}$ and $\brho^{(k)}$ can be
substituted into equation~\ref{est-corlen} to generate estimates
of $\tau$.  The quantiles of these simulated estimates can then be
used to generate a confidence interval.

However, not all the simulated $\bpi^{(i)}$ will define stationary processes
because the true sampling distribution of $\hat\bpi$ is not multivariate
normal.  If the equation:
\begin{equation}
1 = \pi^{(i)}_1 z + \cdots + \pi^{(i)}_p z^p
\end{equation}
has roots inside the unit circle, the resulting AR($p$) process is
not stationary \citep[p.~26]{wei06}.  In this case,
equation~\ref{corlen-ar} does not apply, and the estimate $\hat\tau
= \infty$ should be used instead.  If more than 2.5\% of the simulated
$\bpi^{(i)}$ define nonstationary AR processes, a 95\% confidence
interval will be unbounded; this can be seen in several grid cells of
figure~\ref{fig-results}.

\subsection{Alternatives}

Equation~\ref{corlen-ar} is not the only way to estimate the
autocorrelation time.  One can estimate it with the sample
autocorrelation function evaluated out to some moderate lag chosen
by some rule such as the initial convex sequence \citep{geyer92}.
This method and equation~\ref{corlen-ar} tend to produce similar
estimates.  One advantage of equation~\ref{corlen-ar} is the relative
ease with which confidence intervals may be computed.  A second
alternative is to fit a polynomial model to the log-spectrum of the
sample autocorrelation function \citep{heidelberger81}.  This method
can generate confidence intervals, but has two free parameters and
is difficult to apply to highly correlated chains \citep{plummer10}.
For more information on the estimation of autocorrelation times,
see \citet{act-methods}.

\subsection{Multivariate distributions}

Quite often, some coordinates mix faster than
others.  Consider the linear combination of two coordinates:
\begin{equation}
Y = a X^{(1)} + b X^{(2)}
\end{equation}
If the two coordinates have autocorrelation times $\tau^{(1)}$
and $\tau^{(2)}$, the variance of the mean of $Y$ will be:
\begin{align}
\var(\bar Y_n) &= a^2 \var(\bar X_n^{(1)})
  + 2ab \cov(\bar X_n^{(1)},\bar X_n^{(2)})
  + b^2 \var(\bar X_n^{(2)}) \\
&= a^2 \var(\bar X^{(1)}) + 2ab \sqrt{\var(\bar X^{(1)}) \var(\bar X^{(2)})}
       \corr(\bar X_n^{(1)},\bar X_n^{(2)})
       + b^2 \var(\bar X^{(2)}) \nonumber\\
&\approx
  a^2 \frac{\tau^{(1)}}{n} \var(X^{(1)}) +
  2ab \frac{\sqrt{\tau^{(1)}\tau^{(2)}}}{n}
  \sqrt{\var(X^{(1)}) \var(X^{(2)})} \corr(\bar X_n^{(1)},\bar X_n^{(2)})
  + b^2 \frac{\tau^{(2)}}{n} \var(X^{(2)}) \nonumber
\end{align}
Because the correlation between the two coordinate means is bounded
to $[-1,1]$, if the coordinates have substantially different
autocorrelation times, the first or last term will dominate this
variance.  So, the autocorrelation time of any linear combination
of coordinates will be approximately the autocorrelation time of
the slowest-mixing component.  All comparisons of MCMC methods on
multivariate distributions in this paper take this approach, reporting
figures of merit based on the slowest-mixing component.

The argument of this section only addresses linear combinations of
components.  Nonlinear functions of multiple components---or even
a single component---may have substantially different autocorrelation
times than the autocorrelation times of any linear combination of
components.  The mixing rates of all possible functions cannot be
summarized in a single number; additional information about which
functions are relevant would be necessary.

\section{Methods and distributions for demonstration}
\label{section-example}

Sections~\ref{section-cpu} and~\ref{section-results} both compare
four MCMC methods on four distributions.  Since this paper is
intended to demonstrate methods of comparison rather than advocate
a particular MCMC sampler, I attempted to choose a representative
collection of samplers and distributions.

\subsection{MCMC methods}
\label{methods}

Each MCMC method compared has a single tuning parameter.  The
methods are:
\begin{itemize}
\item Adaptive Metropolis: the adaptive Metropolis--Hastings algorithm
  described by \citet[sec.~2]{roberts09}.  The tuning parameter is the
  standard deviation of the non-adaptive component of the proposal
  distribution, which they fixed at 0.1 (in their equation 2.1).
  This version of Metropolis uses multivariate Gaussian proposals
  with a covariance matrix determined by previous states.  Unlike
  the version of \citeauthor{roberts09}, the implementation used
  here stops adapting after the burn-in period.
\item Univariate Metropolis: Metropolis with transitions that update
  each coordinate sequentially.   The proposals are Gaussian with
  standard deviation equal to the tuning parameter.
\item Shrinking Rank: the shrinking-rank slice sampler described
  by \citet[sec.~5]{thompson10}.  The tuning parameter is $\sigma_c$,
  the initial crumb standard deviation.
\item Step-out Slice: slice sampling with stepping out, updating
  each coordinate sequentially, as described by \citet[sec.~4]{neal03}.
  The tuning parameter is $w$, the initial estimate of slice size.
\end{itemize}

\subsection{Distributions}
\label{dists}

The distributions compared are:

\begin{itemize}

\item $\operatorname{Gamma}(2,1)$: A one dimensional Gamma distribution
  with density proportional to $x e^{-x}$.

\item $N_4(\rho=0.999)$: A four dimensional Gaussian centered at
$(1,2,3,4)$ with covariance matrix:
\begin{equation}
{\bm\Sigma} = \begin{bmatrix} 1 & 0.999 & 0.999 & 0.999 \\
                              0.999 & 1 & 0.999 & 0.999 \\
                              0.999 & 0.999 & 1 & 0.999 \\
                              0.999 & 0.999 & 0.999 & 1 \end{bmatrix}
\end{equation}
This has a condition number of 2870.

\item Eight Schools: A multilevel model in ten dimensions, consisting
of eight group means and hyperparameters for their mean and log-variance
\citep[pp.~138--145]{gelman04}.  Its covariance is well-conditioned.

\item Mixture Ten: A ten-component Gaussian mixture in $\R^{10}$.
Each mode is a spherically symmetric Gaussian with unit variance.
The modes were drawn from a uniform distribution on a hypercube
with edge-length ten.

\end{itemize}

\section{Processor usage and log density evaluations}
\label{section-cpu}
\begin{figure}

\begin{center}\includegraphics[height=7.50in]{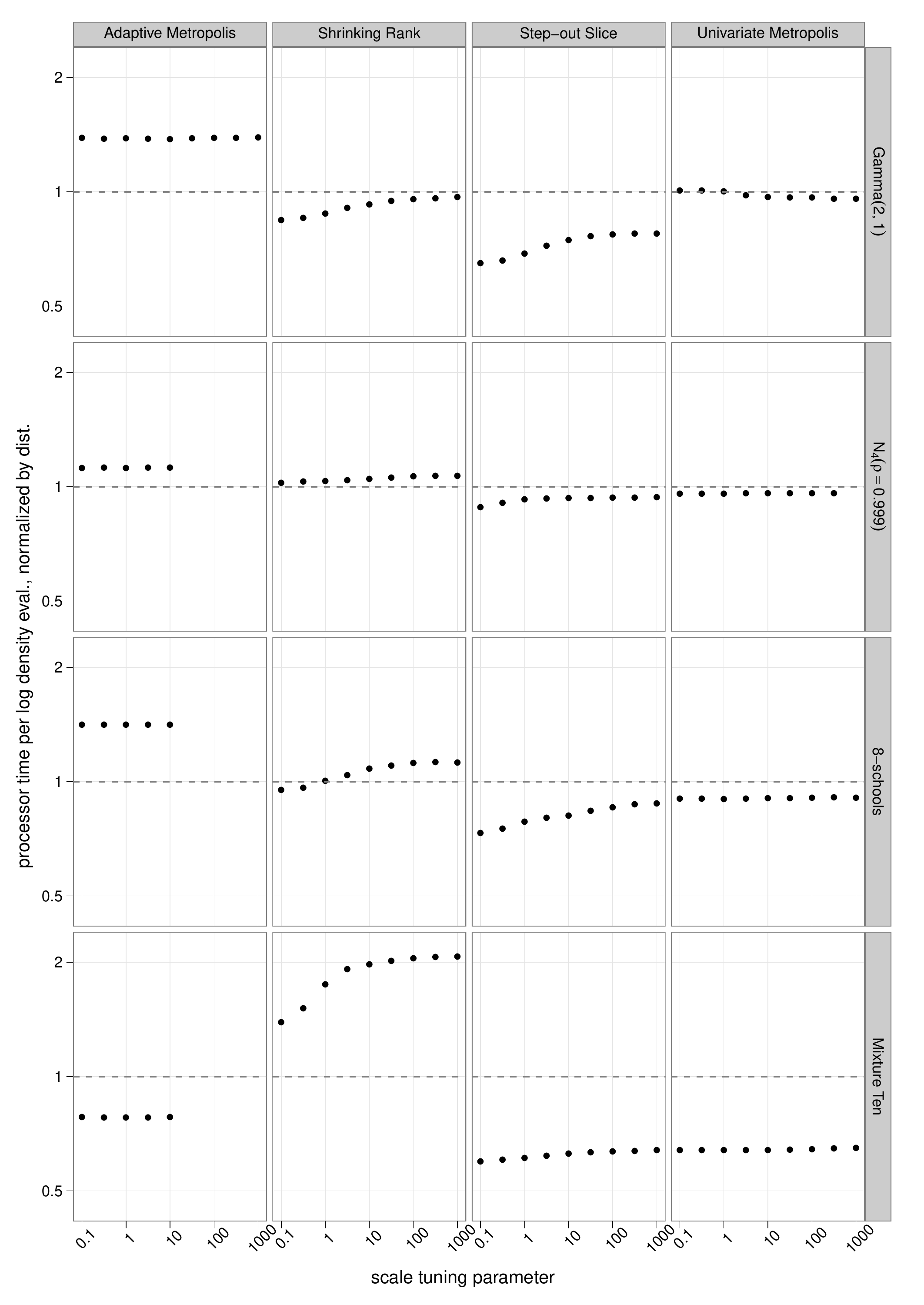}\end{center}
\caption{The ratio of processor usage to log density evaluations,
normalized by distribution.  One can see that this ratio does not
depend much on method choice or tuning parameter; see
section~\ref{section-cpu} for discussion.  Each simulation is of
length 50,000.  The gaps in some of the Adaptive Metropolis plots correspond
to simulations where the autocorrelation time could not be estimated.}
\label{cpu-comparison}
\end{figure}

Before comparing performance of methods on a distribution, we must
choose a figure
of merit.  Here, I use the distributions and methods of
section~\ref{section-example} to justify the choice of log-density
evaluations per MCMC iteration multiplied by autocorrelation time.

If a user's goal is to estimate a parameter to a specific accuracy,
they would want to choose an MCMC method that minimizes processor
time per iteration multiplied by autocorrelation time.  But,
using processor usage on a researcher's machine is problematic
because it depends on irrelevant factors such as the type of machine
the researcher is using, other simultaneous tasks running on the
machine, etc.  Even identically configured simulations may generate
different results.

An alternative to measuring processor usage is counting log-density
function evaluations.  Initialized with the same random seed, an
MCMC simulation that does not use information about the target
distribution other than its log-density function would evaluate
this function the same number of times on machines of
different speeds, on the same machine with different concurrent
processes, and often even when implemented in different programming
languages.  However, this measure does not capture the overhead of
the MCMC method, nor does it capture different ways a distribution
can be represented to the method.  For example, some methods require
gradients to be computed, increasing the processor time per evaluation.
Other methods, like those based on Gibbs sampling, may not compute
log densities at all.

To see the comparability of these two measures, at least on the
distributions and methods described in section~\ref{section-example},
see figure~\ref{cpu-comparison}, which plots the ratio of processor
usage to log-density function evaluations for those samplers and
distributions for a range of tuning parameters.  The ratios are
normalized to the median processor time for a function evaluation
over the simulations of that distribution.  If processor usage and
log density evaluations were perfectly comparable, every plotted
point would lie on the dashed horizontal line indicating a ratio
of one.  For the simulations in figure~\ref{cpu-comparison}, the
measures are at most a small multiple off from each other.

A minor anomaly occurs with the Mixture Ten distribution when simulated
with Shrinking Rank.  Shrinking Rank is the only method that
computes gradients, and the Mixture Ten distribution has gradient
evaluation code that is unusually inefficient relative to its log
density evaluation code.  The ratio does not exceed two, though,
while MCMC method efficiency differences are often several powers
of ten.  Because no significant differences in ratios are observed,
I am inclined to believe that comparing processor usage and comparing
log density evaluations will yield similar conclusions regarding
the merits of various MCMC methods.

\section{Demonstration results}
\label{section-results}

\begin{figure}
\begin{center}\includegraphics[height=7.8in]{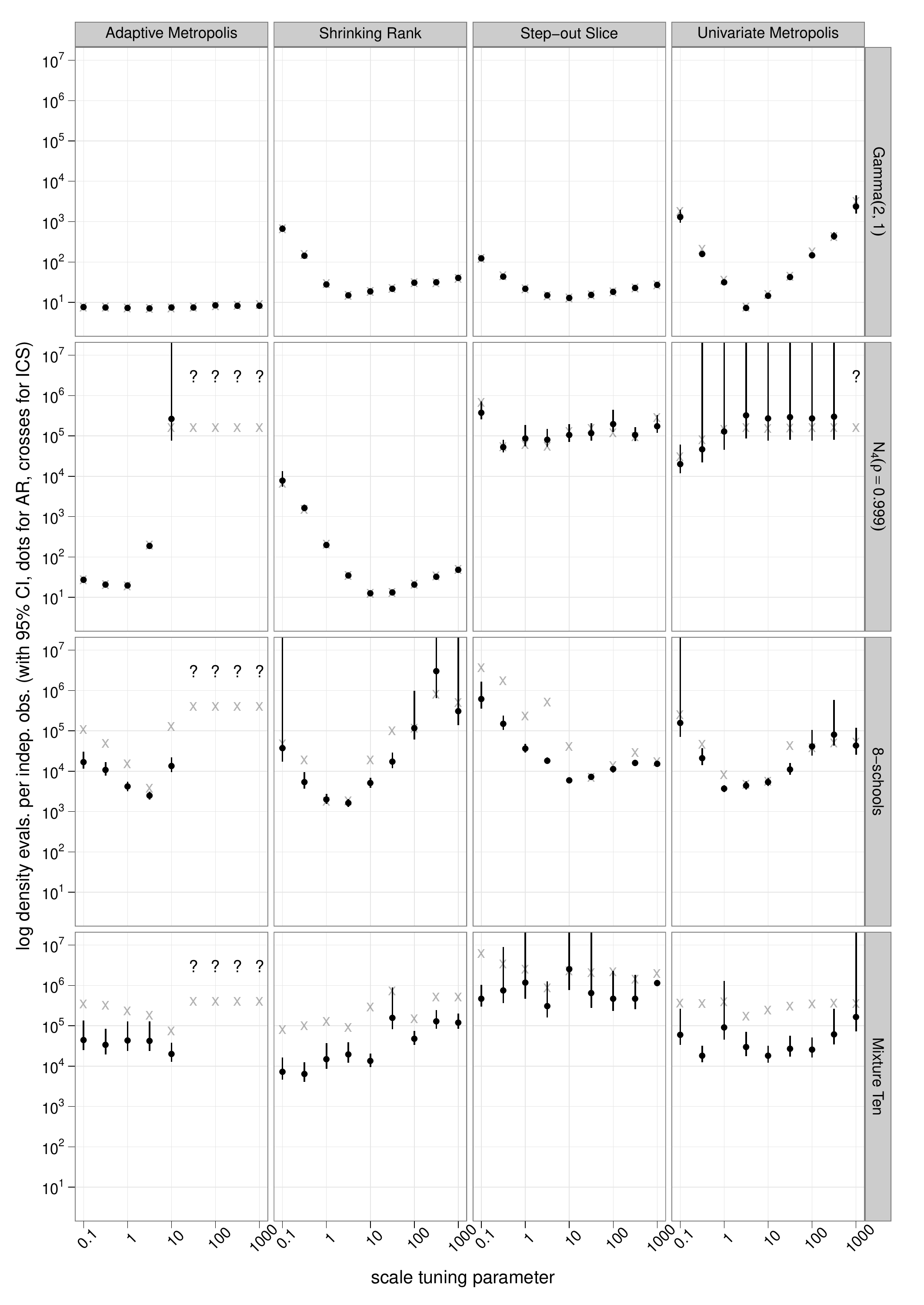}\end{center}
\caption{A demonstration of graphical comparison of performance of
MCMC samplers.  Columns of plots represent samplers; rows represent
distributions.  Each simulation is of length 50,000.  At a glance,
a researcher can see how 144 simulations relate to each other.  See
section~\ref{section-results} for discussion.}
\label{fig-results}
\end{figure}

The figures of merit described in sections~\ref{section-corlen} and
\ref{section-cpu} provide a way of quantifying the performance of
the MCMC methods described in section~\ref{methods} on the distributions
described in section~\ref{dists}.  I simulated each of these
distributions with each sampler for nine values of the sampler's
tuning parameter in the
range $0.1$ to $1000$.  Each chain had a length of 50,000.
Treating the first fifth of each chain as a burn-in period, I
computed the autocorrelation time of each simulation using the
method of section~\ref{section-corlen} and multiplied it by the
average number of log density evaluations the simulation needed per
iteration to generate an estimate of the number of log density
evaluations per independent observation.  Smaller values indicate
better performance.

In figure~\ref{fig-results}, I compare the evaluations per independent
observation by plotting them as dots with a 95\% simulation-based
confidence interval against the tuning parameter for each combination
of distribution and method.  Question marks indicate simulations
with too few unique states to identify an AR model.

The superimposed crosses represent the same figure of merit computed
with the initial convex sequence \citep{geyer92} instead of the AR
model described in section~\ref{estimating}.  The two estimators
mainly differ when the estimated autocorrelation time is close to
or greater than the simulation length, as is the case for the Mixture
Ten simulations.  In this case, they agree inasmuch as both estimators
indicate that the simulation is too short.  Another discrepancy
occurs with Step-out Slice on Eight Schools with tuning parameters
in the range $[0.3,10]$.  These simulations are notable for
occasionally getting stuck briefly in a single part of the state
space.

\subsection{Between-cell comparisons}

Scanning an individual cell in the grid of plots, one can see how
performance varies with the tuning parameter.  Scanning down columns
of the grid of plots, one can see how performance varies from
distribution to distribution for a given MCMC method.   Scanning
across rows of the grid of plots, one can see how performance varies
from method to method for a given distribution.

Doing this, several patterns emerge.  For example, one can see that
Adaptive Metropolis performs consistently well for a variety of
tuning parameters, as long as the tuning parameter is smaller than
the square root of the smallest eigenvalue of the distribution's covariance.
Either it performs well, or does not converge at all.  This suggests
that Adaptive Metropolis users uncertain about the distribution
they are sampling from should err in the direction of small tuning
parameters.

A second pattern is that Univariate Metropolis tends to have U-shaped
comparison plots.  Tuning parameters that are either too small or
too large lead to unacceptable performance.  And, the lowest points
on these plots are still above the flat parts of the Adaptive
Metropolis plots.  On low-dimensional distributions like these,
where the sampler does not know the structure of the target density,
Adaptive Metropolis can be considered to dominate Univariate
Metropolis.

Reading across columns, one can see different sorts of patterns.
For example, the Gamma(2,1) distribution has a relatively flat sequence
of figures of merit across the row of grid cells, suggesting
that sampling from this distribution is not sensitive to choice of
method (if Univariate Metropolis is excluded) or tuning parameter.
The $N_4(\rho=0.999)$ row is also flat within cells, but not between
them, suggesting that for that distribution, one needs to pay
attention to the choice of method, but not as much to the tuning
parameter.

\subsection{Within-cell comparisons}

A different sort of pattern can be seen in slopes of the lines
traced out by the plot symbols.  For Univariate Metropolis and the
shrinking rank method on Gamma(2,1) and Shrinking Rank 
on $N_4(\rho=0.999)$, for tuning parameters less than one, an
increase in the tuning parameter by a factor of ten leads to a
factor of almost one hundred improved performance---a slope of $-2$
in the log domain.  For these tuning parameters the Markov chains
are nearly a one-dimensional random walk; such a random walk will
travel some distance in a number of steps proportional to the
square of the distance \citep[p.~90]{feller68}.

This can be contrasted with the slope of the plot symbols for
Step-out Slice, which is close to $-1$ for small tuning parameters
on the Gamma(2,1) and Eight Schools distributions.  This sampler requires
a number of log density function evaluations linear in the ratio
of slice width to tuning parameter to compute a slice estimate, so
increasing the tuning parameter by some factor improves the performance
by that factor.

A pattern in the slopes for large parameters can be seen with
Univariate Metropolis on Gamma(2,1) with tuning parameters larger
than one.  In those simulations, multiplying the tuning parameter by ten
degrades performance by a factor of approximately ten---a slope of
one in the log domain.  Increasing the tuning parameter by some
factor reduces the acceptance rate by approximately that factor
when the tuning parameter is large.

This behavior contrasts with the two slice samplers, whose figures
of merit follow logarithmic curves for large tuning parameters on
Gamma(2,1).  Each rejected proposal reduces the estimated slice
size by, on average, a factor of two, so an increase in a tuning
parameter above the optimal value by some factor increases the
computation cost by the log of that factor.

None of the patterns of this section would be easily visible if the
results were presented in tabular form.  Figure~\ref{fig-results}
shows summaries of 144 simulations, more than could be absorbed by a
reader if presented as text.

\section{A second demonstration}
\label{second-demo}

\begin{figure}
\begin{center}\includegraphics[height=7.5in]{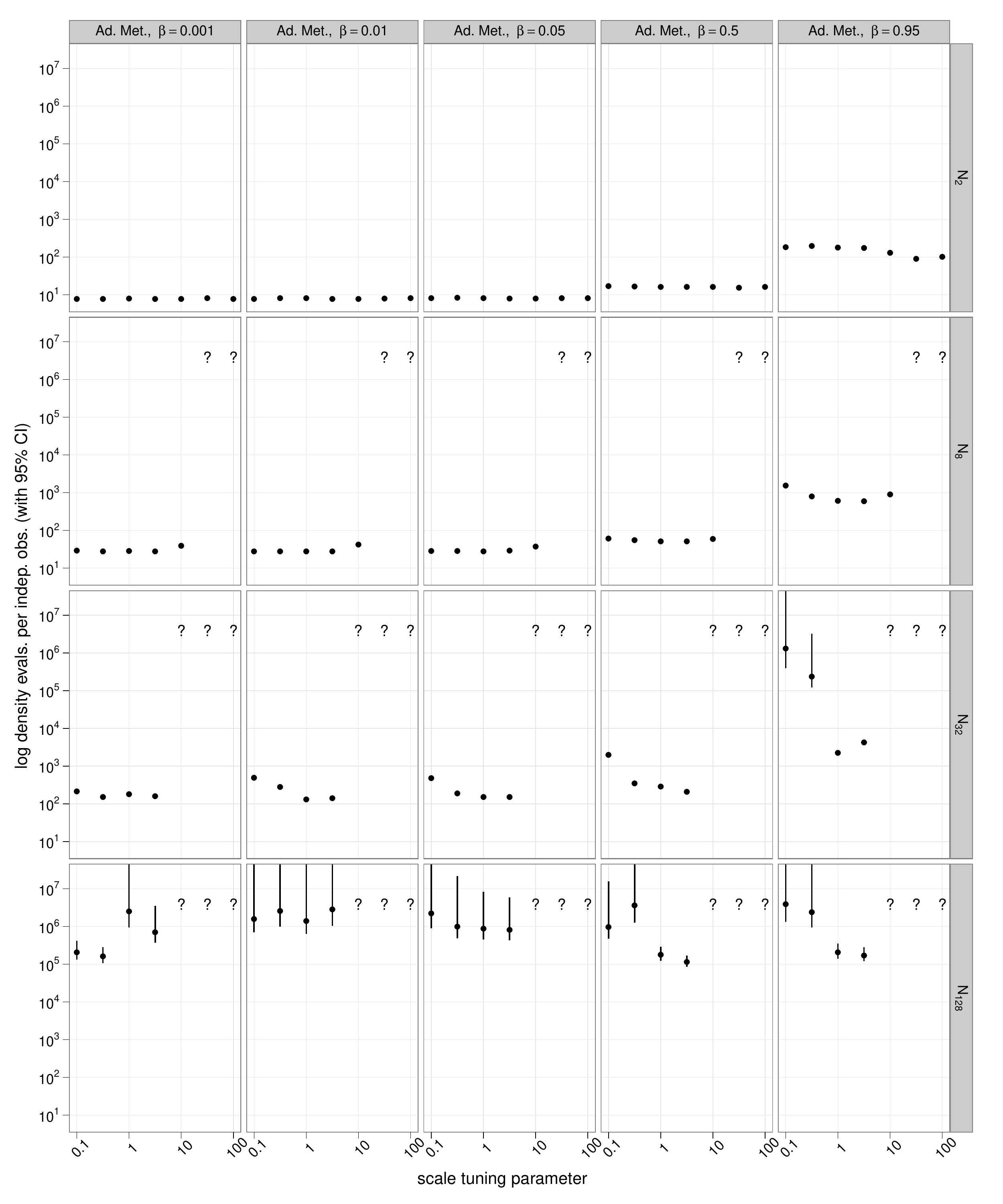}\end{center}
\caption{A demonstration of graphical comparison of performance of
a single sampler, Adaptive Metropolis, with two tuning
parameters.  Columns of plots represent different values of one
tuning parameter, $\beta$; the other is plotted on the horizontal
axis of the subplots.  Rows of plots correspond to increasing
dimensions of badly-scaled Gaussian distributions.   Each simulation
is of length 40,000.  See section~\ref{second-demo} for discussion.}
\label{fig-amh}
\end{figure}

While plots like figure~\ref{fig-results} allow a broad comparison
of distributions and methods, it is often necessary to drill down
further.  The same type of plot can be used when varying extra
parameters associated with either the distribution or the MCMC
method.  Figure~\ref{fig-amh} is an example of such a comparison.
It focuses just on Adaptive Metropolis sampling from badly-scaled
Gaussians.

One tuning parameter to Adaptive Metropolis is $\beta$, the fraction
of proposals drawn from a spherical Gaussian instead of a Gaussian
with a learned covariance.  \citet{roberts09} suggest the value of
0.05.  In this example, I vary it from $0.001$ to $0.95$.

I simultaneously consider the effect of dimensionality on the
performance of the method.  Each target distribution is a multivariate
Gaussian with uncorrelated parameters with variance equal to 1000
in the first coordinate and one in the rest.  The problem dimension
ranges from 2 to 128.

As with the simulations of figure~\ref{fig-results}, Adaptive
Metropolis performs well when the principal tuning parameter is
smaller than the square root of the smallest eigenvalue of the
target distribution's covariance, in this case one.  While the
method performs better in low dimensional spaces, this performance
does not vary much with $\beta$ in any of the dimensions simulated.

\section{Discussion}

Section~\ref{section-results} shows the variation in performance
of MCMC methods while varying method, distribution, and one tuning
parameter.  Section~\ref{second-demo} shows the variation in
performance of a single MCMC method while varying two tuning parameters and
one distribution parameter.  In general, one may use the plots
described in this paper to compare performance of MCMC methods while
varying any three factors.  Comparing performance with tabulated
summary statistics, in contrast, only allows researchers to study
variation in one factor at once, a significant disadvantage.

One might wonder, then, how one could visualize performance as more
than three factors vary simultaneously.  While it may be useful,
computational limits become an obstacle as the number of parameters
increases.  Figure~\ref{fig-results} compares 144 simulations.  Even
if there were a clear way to visualize variation in four or five
factors, one would need to run thousands of simulations to make
such a plot.  The plots of this paper display approximately as much
information as is currently feasible to gather.

Other variations on these plots are possible.  The most common is
to replace confidence intervals by multiple plot symbols, each
representing a replication of the same simulation with a different
random seed, so that multiple symbols are plotted in a column for
a given tuning parameter.  In this case, making the symbols partially
transparent makes the graph easier to read.

If one is willing to forego all measures of uncertainty under
replication, one can omit the confidence intervals and plot figures
of merit for multiple values of a second factor, such as a second
tuning parameter, in a single grid cell, differentiating levels of
this factor by color or plot symbol.  However, this technique can
produce confusing graphs if the factor represented by color or plot
symbol is not naturally nested in the factor represented by columns
of plots, so one should not vary a factor like problem dimension
inside a single grid cell.

\section{Software}

SamplerCompare is an R package that allows researchers to run
simulations and generate plots like those described in this paper.
It supports MCMC methods and distributions implemented in R and C
and is available on CRAN at
\url{http://cran.r-project.org/web/packages/SamplerCompare}.

\section{Acknowledgments}

I would like to thank Radford M. Neal for his extensive comments
on the draft manuscript.

\bibliographystyle{apalike}
\bibliography{corlen}

\end{document}